# Thermal Modulation of Gigahertz Surface Acoustic Waves on Lithium Niobate

Linbo Shao[1,2,*], Sophie W. Ding[1], Yunwei Ma[2,3], Yuhao Zhang[2,3], Neil Sinclair[1,4], Marko Lončar[1,*]

[1]John A. Paulson School of Engineering and Applied Sciences, Harvard University, Cambridge, Massachusetts 02138, USA

[2]Bradley Department of Electrical and Computer Engineering, Virginia Tech, Blacksburg, Virginia 24061, USA

[3]Center for Power Electronics Systems (CPES), Virginia Tech, Blacksburg, Virginia 24061, USA

[4]Division of Physics, Mathematics and Astronomy, and Alliance for Quantum Technologies (AQT), California Institute of Technology, Pasadena, California 91125, USA

*Correspondence to: shaolb@vt.edu (L.S.); loncar@seas.harvard.edu (M.L.)

**Abstract**

Surface acoustic wave (SAW) devices have wide range of applications in microwave signal processing. Microwave SAW components benefit from higher quality factors and much smaller crosstalk when compared to their electromagnetic counterparts. Efficient routing and modulation of SAWs are essential for building large-scale and versatile acoustic-wave circuits. Here, we demonstrate integrated thermo-acoustic modulators using two SAW platforms: bulk lithium niobate and thin-film lithium niobate on sapphire. In both approaches, the gigahertz-frequency SAWs are routed by integrated acoustic waveguides while on-chip microheaters are used to locally change the temperature and thus control the phase of SAW. Using this approach, we achieved phase changes of over 720 degrees with the responsibility of 2.6 deg/mW for bulk lithium niobate and 0.52 deg/mW for lithium niobate on sapphire. Furthermore, we demonstrated amplitude modulation of SAWs using acoustic Mach Zehnder interferometers. Our thermo-acoustic modulators can enable reconfigurable acoustic signal processing for next generation wireless communications and microwave systems.

**Introduction**

Surface acoustic wave (SAW) devices [1-4] have been widely used in microwave signal processing to realize oscillators, microwave filters, and delay lines. This is since SAW propagates at a velocity five orders of magnitude smaller than that of electromagnetic wave at the same frequency, and thus SAW devices have much smaller wavelengths than their electromagnetic counterparts. This enables high density integration and reduced crosstalk between SAW devices, which has been leveraged to achieve efficient compression, correlation, and Fourier transform of radio frequency (RF) pulses, for example, in a compact footprint. Acoustic-wave devices have successfully pushed their operating frequency to tens of gigahertz [5-9] and thus have become a promising platform for microwave signal processing for next generation wireless communications.

SAW has also served as a versatile interface between different quantum bits (qubits), including electron spins [10-12], superconducting qubits [13-20], and optical photons [20-25]. SAW devices utilize piezoelectric materials, such as ST-X quartz [14], aluminum nitride (AlN) [13,20,26-29], scandium-doped aluminum nitride (Sc-AlN) [30-32], gallium arsenide (GaAs) [18,19], gallium nitride (GaN) [33], zinc oxide (ZnO) [34], lithium tantalate [35,36], and lithium niobate (LN) [15-17,21,22,37-42], to convert signals between electromagnetic waves and acoustic waves. To date, most acoustic devices are linear, passive, and reciprocal components, which limit their functionality. Several efforts have been made to develop nonlinear, active, and nonreciprocal SAW devices. For example, nonreciprocal acoustic devices have been demonstrated by circulating fluids [43], spatiotemporal modulations [44,45], acoustoelectric coupling in semiconductors [46,47], magnetoelastic coupling [48] and non-Hermitian physics [49-51]. Tunable SAW devices have been achieved using mechanical nonlinear deformations [52], piezoelectrically-actuated deformation [53], and electro-acoustic effects [45] but still require high voltages to achieve a full $\pi$ phase shift.

Here, we demonstrate stable and efficient modulation of acoustic waves propagating on both bulk and thin-film LN on sapphire (LNSa) [40] substrates. Our approach leverages temperature dependence of LN elasticity [54] to control the phase velocity of acoustic waves by tuning the temperature of the waveguide that the waves propagate in. This is accomplished using microheaters located close to the acoustic waveguides (Fig. 1). Our thermo-acoustic modulators provide an efficient and stable approach to control the SAWs on-chip, paving the way for more versatile acoustic-wave microwave signal processing circuits.

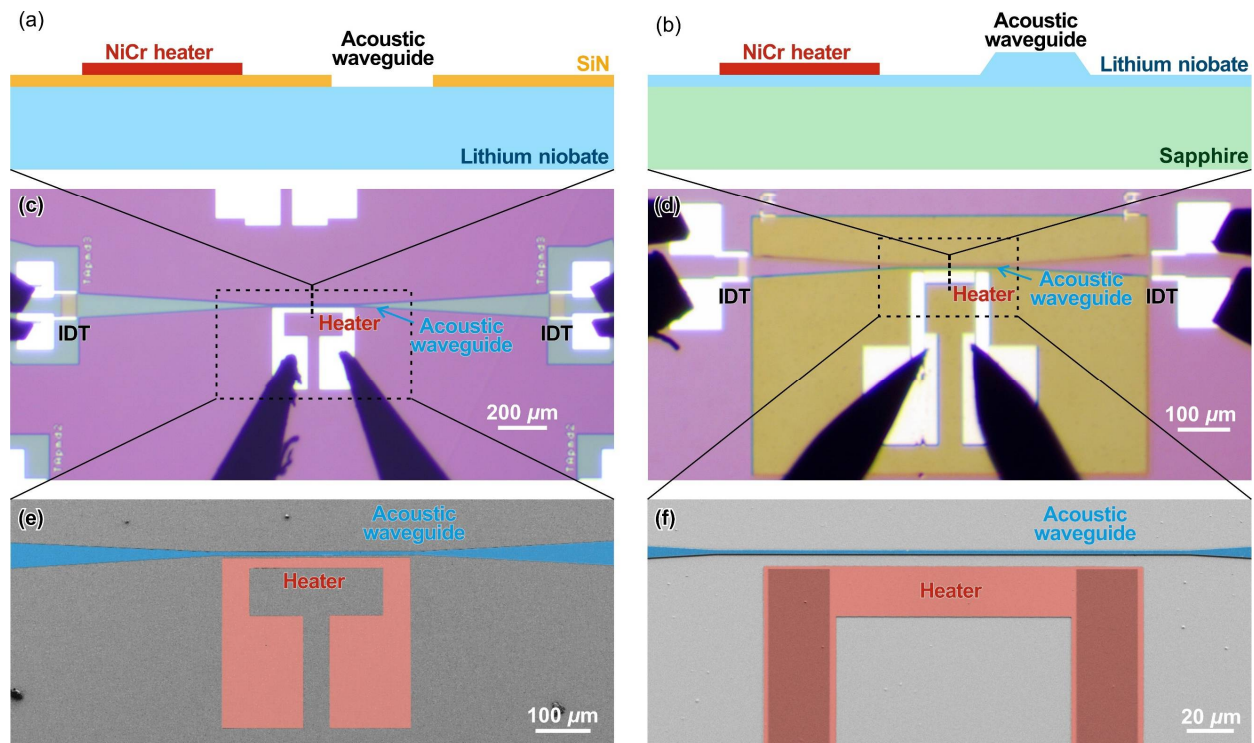

**Fig 1. Integrated thermo-acoustic modulators based on the bulk lithium niobate (LN) substrate in (a)(c)(e) and the thin-film LN on sapphire (LNSa) in (b)(d)(f).** (a)(b) Illustrations of the cross sections of the thermo-acoustic modulators. Nanofabricated microheaters made of nickel-chromium (NiCr) alloy are used to tune the temperature of the acoustic waveguides defined by (a) an opening (slot) made of silicon nitride (SiN) film or (b) by etching lithium niobate (LN) film (c)(d) Optical microscopy images of the thermo-acoustic modulators. Interdigital transducers (IDTs) are used to excite and detect the acoustic waves. Microwave ground-signal (GS) probes are used to contact the IDT electrodes, and low frequency probes are used to deliver current to the microheaters. The dashed boxes indicate the zoom-in regions shown in (e)(f). (e)(f) False colored scanning electron microscopy (SEM) images of the thermo-acoustic modulators.

**Design and Fabrication of Thermo-acoustic Modulators**

Our thermo-acoustic modulators include acoustic waveguides, microheaters, and interdigital transducers (IDTs). Acoustic waves are generated piezoelectrically by the IDT, tapered into the acoustic waveguide, tapered-out to another IDT region where they are detected (Figs. 1(c) and 1(d)). The microheater is situated adjacent to the waveguide.

For bulk LN platform, we use a 128° Y-cut black LN substrate (the acoustic waves propagate in the crystal X direction) which allows efficient IDTs on the bulk substrate. The electromechanical coupling efficiency $k^2$ for the IDT is about 5.5%. As a demonstration, the acoustic wavelength is set to 2.8 $\mu$m at the IDT region for the bulk LN platform. The acoustic waveguides are defined by opening a slot in the overlay silicon nitride (SiN) thin film [45] (Fig. 1(a)). Since SiN is a harder (larger stiffness coefficients) material, it allows a larger acoustic velocity than LN. Thus, a 7-$\mu$m-wide slot in the SiN thin film will create a region with lower acoustic velocity and guide the acoustic waves. The SiN thin film is deposited by plasma-enhanced chemical vapor deposition (PECVD) and patterned by photolithography using a laser direct writer and reactive-ion etching (RIE) using carbon tetrafluoride ($CF_4$) and trifluoromethane ($CHF_3$) gases. The thickness of SiN is about 500 nm, which allows efficient guiding of gigahertz acoustic waves and feasibility in fabrication.

For the LNSa platform, we use a 620-nm-thick X-cut LN on a sapphire substrate, and the acoustic wave propagates in the crystal Y direction of LN. The acoustic wavelength is chosen to be 1.7 $\mu$m at the IDT region, which yields the optimal electromechanical coupling efficiency $k^2$ as high as 20% for the shear mode [40]. The pitch of the IDT is half of the acoustic wavelength. The acoustic rib waveguide is 7 $\mu$m in top width and etched down by 550 nm by RIE using argon gas (Fig. 1(b)). The sidewall angle is about 70°. The acoustic waveguide supports the fundamental Rayleigh and shear modes (Fig. 2(a)). The propagation losses of acoustic waveguides have been previously investigated for both the bulk LN [45] and LNSa platforms [40].

On both platforms, aluminum (Al) IDTs are patterned by a lift-off process with thickness of 120 nm (100 nm) for the bulk LN (LNSa) devices. Al is deposited by electron-beam evaporation or thermal evaporation, and we find no performance differences between the two deposition processes. The microheaters are defined by another layer of lift-off process with nickel/chromium (NiCr, 80/20 wt%, electrical resistivity $\rho = 1.9 \times 10^{-6}$ $\Omega \cdot$m) using electron-beam evaporation. We design microheaters with dimensions of 10 to 25 $\mu$m in width, 100 to 250 $\mu$m in length, and 100 to 200 nm in thickness, which result in electric resistance from 40 to 500 $\Omega$ for different devices and applications. To reduce the contact resistance between the microheater and its contact pads, an additional 100-nm-thick Al layer is deposited on the NiCr layer, as shown by the darker area in Fig. 1(f). The microheaters are placed with a minimum edge-to-edge clearance of 6 $\mu$m to ensure that the microheater does not cause additional propagation loss to the acoustic waves.

**Simulation of Thermo-acoustic Modulation**

We conduct numerical simulations to investigate the performance of the thermo-acoustic modulators. We simulate the waveguide modes and the thermal response of the devices using finite element method (FEM) simulations in COMSOL Multiphysics (Fig. 2). For the simulation of the guided acoustic modes, Bloch boundary conditions are imposed for the waveguide, so that the eigenmodes correspond to the guided modes with the same wavelength. The modes with distinct displacements at the lowest frequencies are the fundamental modes. There are two types of modes in the LNSa waveguide, the shearing and Rayleigh modes (Fig. 2(a)), at 2.7 GHz and 2.9 GHz respectively, with an acoustic wavelength of 1.4 $\mu$m in the waveguide. Note that in the IDT region, the acoustic wavelength is 1.7 $\mu$m at the same frequency, which is longer than that in the waveguide; this is due to the larger acoustic velocity at the IDT region than that at the waveguide. For the bulk LN waveguides, the relevant mode is a Rayleigh mode at 1.4 GHz with an acoustic wavelength of 2.9 $\mu$m (Fig. 2(a)).

The thermal response of the modulator is simulated at a cross-section of the device perpendicular to the SAW propagation direction (Fig. 2(b)). The device is simulated in quasi-3D at room temperature (293.15 K) and atmosphere with periodic boundary conditions applied to the surfaces of the cross-sectional slice. The fixed temperature boundary condition is applied to the other surfaces of the domain. We perform simulations with different domain sizes to ensure to ensure the validity of our results. The heater electrode is modeled as the heat source, and heat propagates through the materials and brings the waveguide to a higher temperature in the steady-state solution (Fig. 2 (b)).

To study the frequency domain behavior of the bulk LN and LNSa modulator, we modulate the power applied to the electrode, which can be treated as a perturbation to the steady-state solution. We simulate the waveguide temperature variation as a function of the modulation frequency. The 3-dB cutoff frequency for thermal modulation is 8 Hz (20 Hz) for the bulk LN (LNSa) modulator (Fig. 2(d)). The difference in the frequency response of the two modulators reflects the difference in the material thermal conductivity, as sapphire is more thermally conductive than bulk LN.

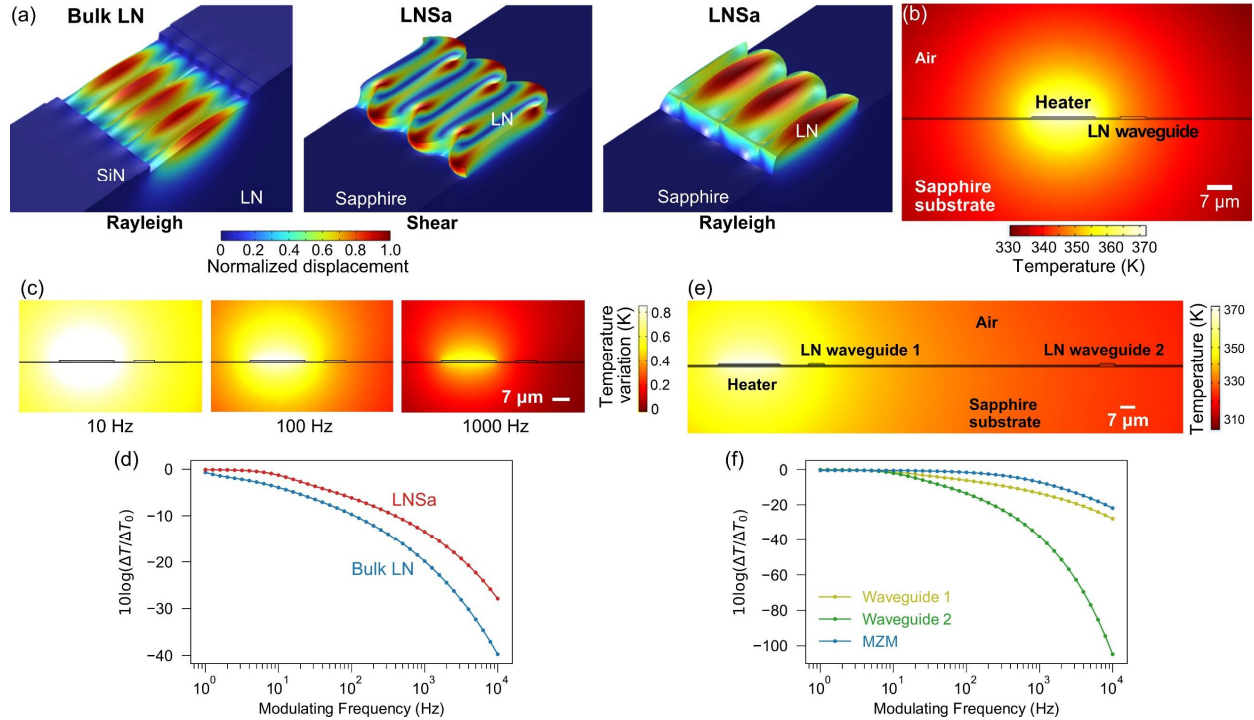

**Fig. 2 Simulation of thermo-acoustic modulators.** (a) Acoustic waveguide simulation of the acoustic modes in the bulk LN and LNSa modulator. (b) Steady-state thermal response of the LNSa modulator: the relative location and the size of the waveguide and electrode are consistent with the fabricated device. The heatmap indicates the temperature. (c) Perturbative solutions at specific modulation frequencies for the LNSa modulator: as the frequency increases, the thermal response, and therefore the phase modulation of the waveguide, is suppressed. (d) Frequency response of thermally driven bulk LN and LNSa phase modulators. (e) The cross-section geometry of the Mach-Zehnder modulator (MZM) and the heatmap of the steady-state solution. (f) Frequency response of thermally driven LNSa MZM. Phase modulation of each arm is also shown.

To achieve amplitude modulation of acoustic waves, we demonstrate a LNSa thermo-acoustic Mach-Zehnder modulator (MZM). The MZM utilizes the phase difference between two arms (waveguides) to achieve amplitude modulation via interference when two arms are merged at the output. To construct thermo-acoustic MZMs, we place the microheater at different distances away from the two acoustic waveguides, and thus the SAW in each arm accumulates different phases. Intuitively, the output of the MZM depends on the temperature difference between two waveguides. This is different from the phase modulator, in which the temperature of individual waveguides compared to the environmental temperature determines the phase shift. Therefore, the MZM shows a 3-dB modulation bandwidth of about 400 Hz, which is much higher than that of the phase modulator (Fig. 2(f)). The microheater is placed 6 $\mu$m away from the waveguide in one arm, but 106 $\mu$m away from the waveguide in the other.

## Experimental Characterization of Thermo-acoustic Phase Modulators

We experimentally measure the $S$ parameters of the thermo-acoustic phase modulators using a vector network analyzer (VNA), while controlling the power applied to the microheater (Fig 3(a)). The IDTs are connected by microwave ground-signal (GS) probes (GGB 40A), and the microheaters are contacted by low-frequency probes. The amplitude and phase of $S$ parameters are calibrated to the probes. We control the phase of the output acoustic waves by tuning the power delivered on the microheaters (Fig. 3(b)). The bulk LN modulator exhibits a phase response of –2.6 deg/mW, compared to only –0.52 deg/mW for the LNSa modulator. The phase shift is measured at the optimal frequencies with largest transmission $S_{21}$. The response of the phase shift to the applied microheater power is linear.

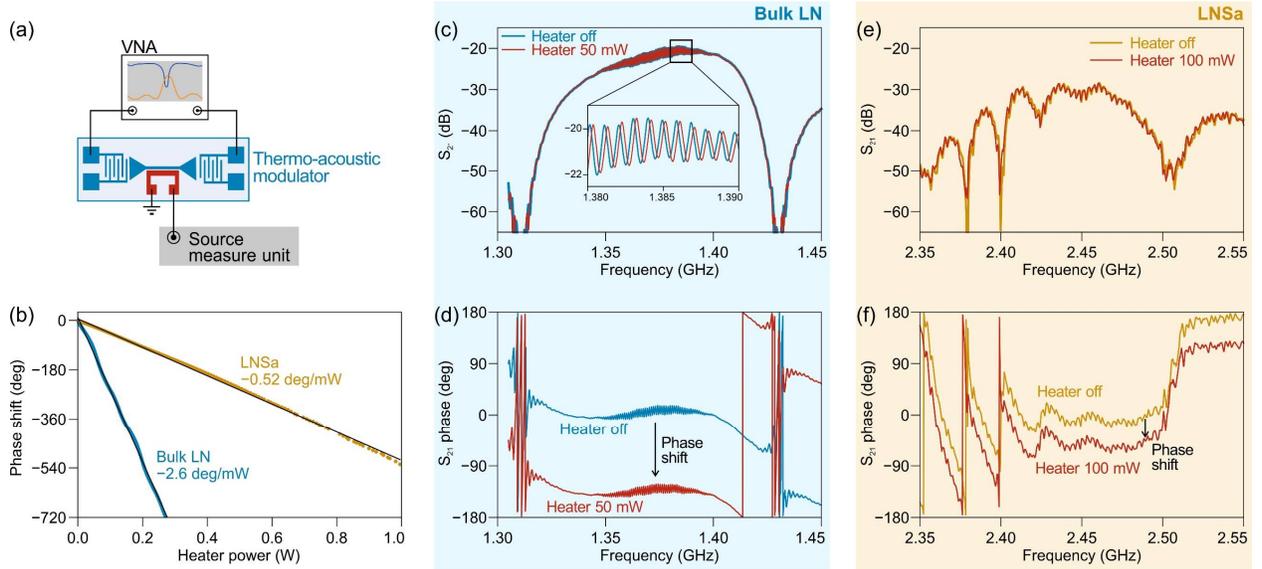

**Fig. 3 Experimental characterization of the thermo-acoustic phase modulators on bulk lithium niobate (LN) and LN on sapphire (LNSa).** (a) Experimental setup. A vector network analyzer (VNA) measures the $S$ parameter of the modulator, and a source measure unit applies voltage on the micro heater while measuring the current. (b) Phase shift for various powers applied to the heater. The phase shift is measured at SAW frequencies of 1.385 GHz and 2.46 GHz in the case of the bulk LN and LNSa modulator, respectively. We speculate that the periodic variations seen in the case of bulk LN device are caused by the instability of the source used to control the heater power. (c-f) Transmission measurement with heaters off and on for (c)(d) the bulk LN modulator and (e)(f) the LNSa modulator. (c)(e) The amplitudes of $S_{21}$ are not affected by the heater. (d)(f) The phases of $S_{21}$ are shifted by the heater. The linear frequency-phase responses of $S_{21}$ due to the acoustic-wave group delays are subtracted in (d)(f).

We measure the transmission $S_{21}$ spectra of our modulators when the microheater is switched on and off (Figs. 3(c)-3(f)). The acoustic-wave bandwidths of our thermo-acoustic modulators are determined by their IDTs. The acoustic-wave bandwidth is about 90 MHz (from 1.33 to 1.42 GHz) for the bulk LN modulator (Fig. 3(c)) and about 70 MHz (from 2.41 to 2.48 GHz) for the LNSa modulator (Fig. 3(e)). As expected for phase modulators, the amplitude of $S_{21}$ does not change when the heater is switched on (Figs. 3(c) and 3(e)), while the phase of $S_{21}$ is shifted (Figs. 3(d) and 3(f)). The $S_{21}$ phases in Figs. 3(d) and 3(f) are offset by the delay time at the central frequencies of each device. For the LNSa modulator, we note that different types of acoustic waves might be excited by the IDT simultaneously, and thus lead to different dispersions (delay times) and interference in the 2.35 to 2.40 GHz frequency range. Overall, our thermo-acoustic modulator can provide a phase shift over frequencies within their IDT bandwidths.

## Frequency response of the thermo-acoustic modulators

We now characterize the frequency response of our thermo-acoustic modulators. A function generator provides small signals at different frequencies and DC offsets to the microheater (Fig. 4(a)). A microwave signal generator provides continuous signals at the frequencies with maximum transmission; 1.385 GHz for the bulk LN modulator and 2.44 GHz for the LNSa modulator. A commercial in-phase quadrature (IQ) mixer is used to downconvert the output SAW signals to low frequency in-phase (I) and quadrature (Q) signals (Fig. 4(b)), which are captured by an oscilloscope and converted to the amplitude and phase of the output acoustic waves (Fig. 4(c)). Ideally, the phase modulators do not change the amplitude of the SAW. The observed small variation in the amplitude of the output acoustic waves might be due to the interference between different types of acoustic waves or misalignment of I/Q signals.

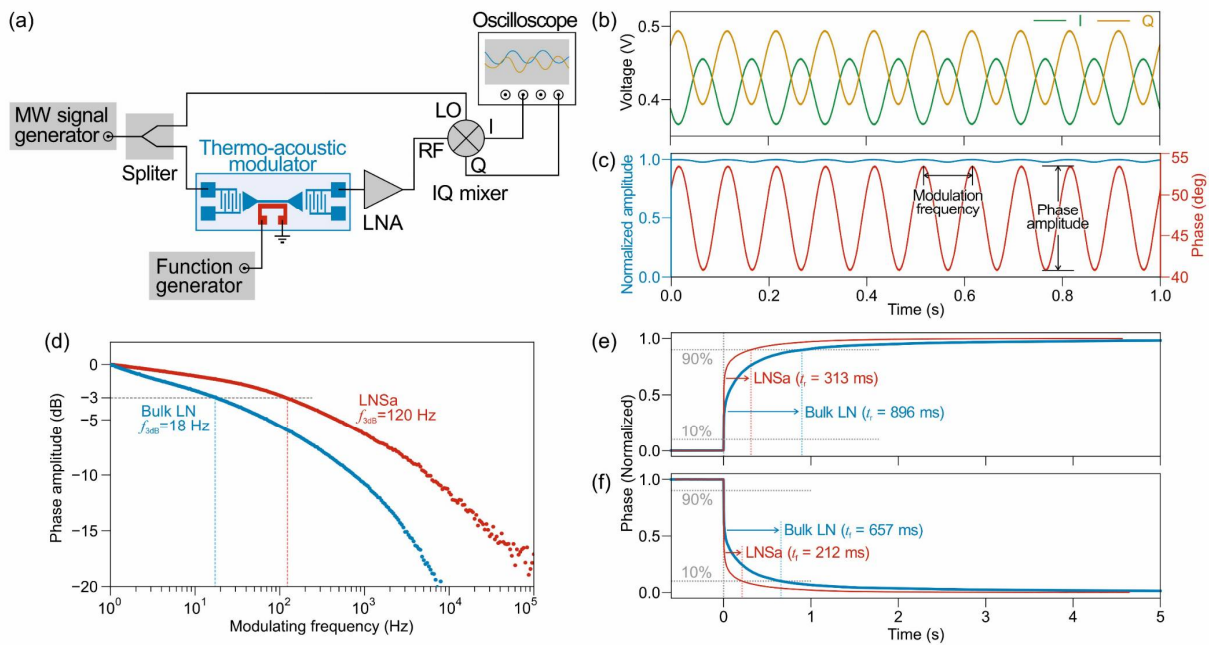

**Fig. 4 Characterization of the frequency and transient response of thermo-acoustic phase modulators.** (a) Experimental setup. The microwave (MW) signal generator provides a source signal at 1.385 GHz for the bulk LN modulator and 2.44 GHz for the LNSa modulator. The source signal is split into two paths, one connects the input IDT of the device, and the other one feeds the local oscillator (LO) input of an IQ mixer. A function generator applies a time varying signal on the microheater. The modulated signal from the output IDT is amplified by a low-noise amplifier (LNA) and downconverted by the IQ mixer. The IQ signals are finally measured by an oscilloscope. (b) Measured I/Q signals and (c) calculated amplitude and phase of the LNSa modulator output when a 10-Hz sine wave with amplitude of 0.3 V and offset of 3.5 V is applied to the heater. (d) The frequency response of the phase amplitude for the bulk LN and the LNSa modulator. (e)(f) Transient response of the output phase when the heater is (e) turned on and (f) turned off by a square wave. The step voltage is set to 3 V for the bulk LN modulator and 3.5 V for the LNSa modulator to achieve phase shifts about 120 deg.

The measured frequency responses show a 3-dB bandwidth of 18 Hz for the bulk LN modulator and a higher bandwidth of 120 Hz for the LNSa modulator (Fig. 4(d)). We speculate that the differences between the measured results and numerical simulations are due to following factors: (1) our simulation is quasi-3D and does not take the length of the heater in the account, while fabricated microheater has a limited length; (2) the material properties that are used in simulations may not be accurate for the black LN and the LN thin films.

We also characterize the transient response for large signals applied to the microheaters. We apply a step function with a low voltage of 0 V and a high voltage of 3 V (3.5 V) for the bulk LN (LNSa) modulator. We observe a rise time of 896 ms (313 ms) when turning on the microwave heater for the bulk LN (LNSa) modulator and a fall time of 657 ms (212 ms) when switching off. The transient response times for large signals are longer than the transient times converted from their 3-dB bandwidths. The difference between the rise and fall times reflects the different time constants in the heating and cooling procedures, which are not symmetric. For the heating, the acoustic waveguide is heated by the nearby microheater, while for the cooling, the heat is dissipated to the environment. We also note that the transient curves shown in Figs. 4(e) and 4(f) are not exponential due to the complex thermal response of our thin-film stacks.

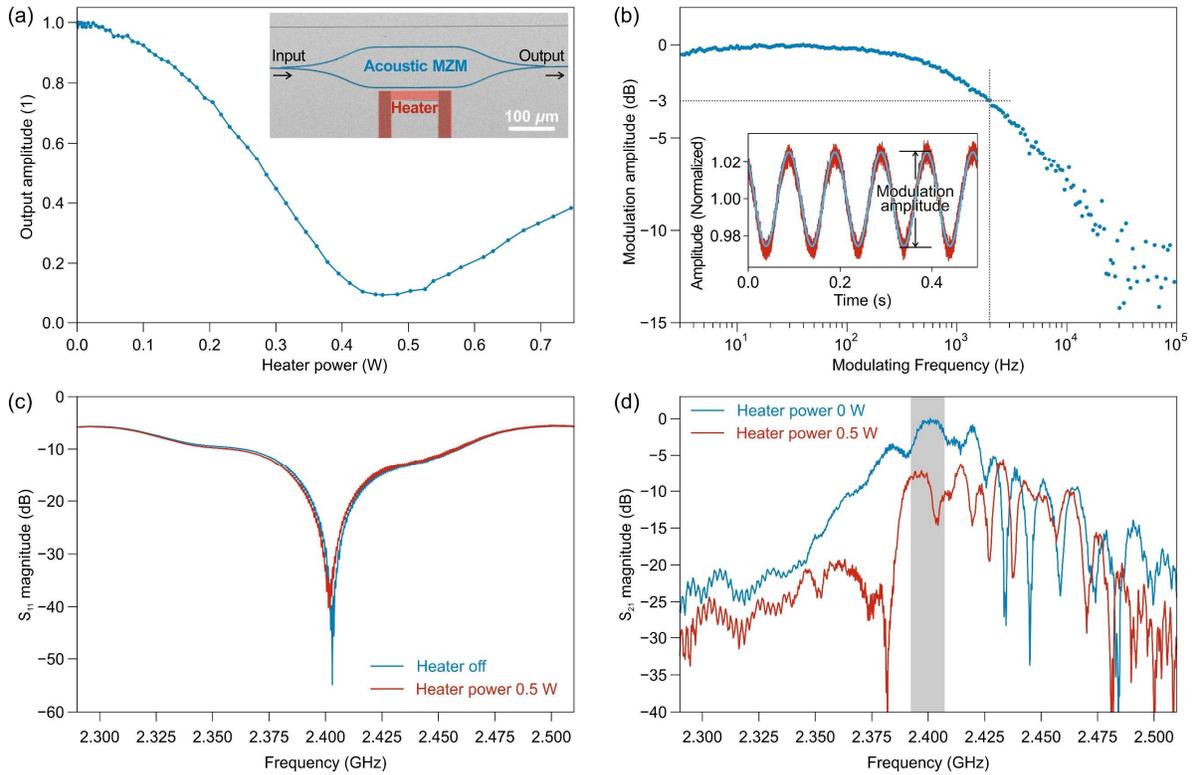

**Fig. 5 Characterization of thermo-acoustic Mach-Zehnder modulator (MZM)**. (a) Amplitude of the output acoustic wave for varied heater power. Inset: SEM image of a thermo-acoustic MZM. (b) The frequency response of the MZM. Inset: Modulated amplitude of the output acoustic waves. (a)(b) Same setup as shown in Fig. 4(a) is used to characterize the device, and the amplitude is calculated from the I/Q signals captured by the oscilloscope. (c) $S_{11}$ and (d) $S_{21}$ parameter measurements of the thermo-acoustic MZM with the heater off and at 0.5 W. Same setup as shown in Fig. 3(a) is used to measure the S parameters in (c)(d). The thermo-acoustic MZMs are fabricated on the LN on sapphire (LNSa) platform. Devices used in (a)(b) and (c)(d) are of the same design but from different fabrication batches, and thus the central frequencies are slightly different. The central frequencies for the device in (a)(b) is 2.45 GHz and that in (c)(d) is 2.40 GHz.

**Thermo-acoustic Mach-Zehnder Modulator**

We use LNSa platform to demonstrate the thermo-acoustic MZM (Fig. 5(a) Inset). Using the same setup shown in Fig. 4(a), we measure the output acoustic amplitude at various heater powers (Fig. 5(a)). For an input acoustic frequency of 2.45 GHz. The minimum output amplitude is observed at a heater power of 0.45 W, and the minimum output power is about 10% of the maximum output power when the heater is off, resulting in an extinction ratio of ~10 dB. We further measure the frequency response of the thermo-acoustic

MZM. The heater is biased at about 0.2 W, and small signals with various frequencies are applied to the heater. The modulation amplitude is obtained from the captured I/Q signals (Fig. 5(b)). Here, we observe the 3 dB bandwidth of 2.0 kHz for the thermo-acoustic MZI modulator, which is much larger than that of the phase modulator, as expected. Generally, a smaller lateral distance between two MZM arms will result in a larger modulation bandwidth. However, reducing the lateral distance between two arms will reduce the temperature difference and thus reduce the modulation amplitude at low frequency.

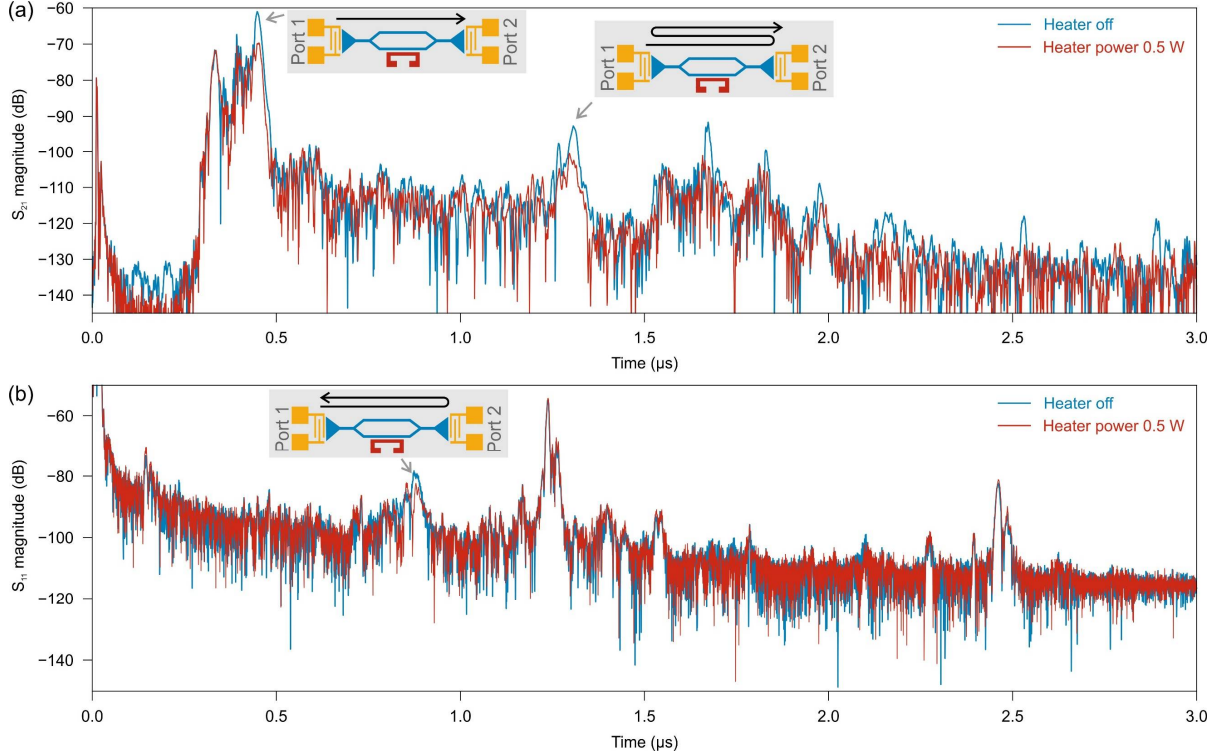

**Fig. 6 Time-domain analysis of the thermo-acoustic MZM.** The multiple peaks observed in (a) the transmission and (b) reflection signals indicate possible involvement of multiple acoustic modes and reflections. The single pass traveling time for the main acoustic mode is about 450 ns. The central frequency for time domain analysis is 2.40 GHz. The corresponding *S*-parameters are shown in Figs. 5(c) and 5(d).

The limited extinction ratio (~ 10 dB) of the amplitude modulation may be due to (1) uneven power splitting between two MZM arms or (2) coupling to higher-order acoustic modes, which have different propagation constants. Comparing the S-parameter spectra with the heater off and on (Figs. 5(c) and 5(d)), we observe that the extinction ratio is frequency dependent. The interference patterns between 2.42 to 2.45 GHz in Fig. 5(d) indicates possible involvement of multiple acoustic modes in our MZM. The $S_{11}$ spectra are mainly defined by the IDTs, and we observe good impedance matching at the central frequency. Furthermore, we have performed time-domain measurements and analysis (Keysight P5004A) of the acoustic MZM (Fig. 6). The multiple peaks in the transmission ($S_{21}$) and reflection ($S_{11}$) signals also infer the existence of multiple modes and reflections. The single pass travel time of the acoustic wave is about 450 ns. In addition, we optically imaged the profiles of guided acoustic waves and observed patterns supporting mixing with higher-order modes (data not shown). Future developments and optimizations are needed to improve the extinction ratio of acoustic MZMs for practical applications.

## Stability of the Thermo-acoustic Phase Modulator

Stability over time is one of the critical specifications for SAW devices. To characterize the stability, we measure the output of our thermo-acoustic phase modulator over different time frames of 100 seconds and one hour (Fig. 7). The heater is driven at a power of 0.2 W. We observe a variation in phase within a 0.15-degree (0.5-degree) range over 100 seconds (one hour) (Fig. 7). We speculate that the remaining phase variations are due to the temperature fluctuation of the environment. We note that thermally controlled LN modulators feature a better stability over a long time than our previous modulators based on electrical fields [45].

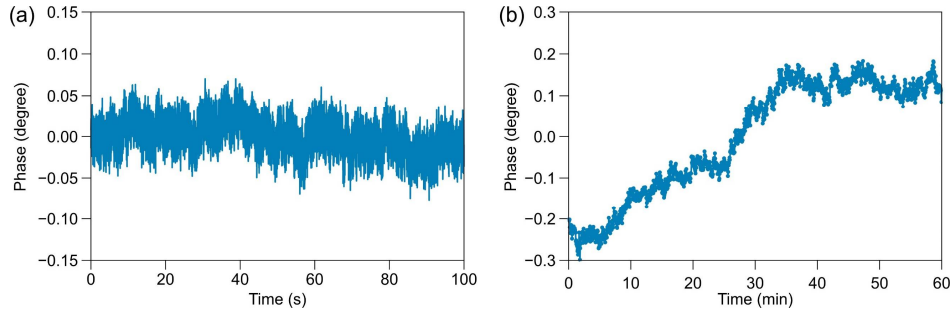

**Fig. 7 Characterization of the stability of the thermo-acoustic phase modulators.** The phase of the acoustic signal after the phase modulators over (a) 100 seconds and (b) one hour. The output phase is calculated from the I/Q signals, as shown in Fig. 4(a). The sampling rate of the phase is 1000 samples per sec (sps) in (a). For each data point in (b), the capture time is 1 second with a sampling rate of 1000 sps, and the phase is captured at 3 second intervals, allowing data transfer from the oscilloscope.

## Discussion and Outlook

We demonstrate thermal modulation of GHz acoustic waves propagating two platforms – bulk LN and LNSa. As the thermal conductivity of LN is lower than that of sapphire, the bulk LN modulators feature larger phase shift per unit heater power than LNSa modulators, while LNSa modulators show a larger frequency bandwidth. For the integration of other LN devices on a single chip, such as optical waveguides and modulators, rib waveguides fabricated using thin-film LN on sapphire can simultaneously guide the acoustic waves and optical light, while the bulk LN platform may require an overlay high refractive index material, such as silicon, or an ion-diffused region to guide the optical light. Compared to the electro-acoustic modulators [45], our thermo-acoustic modulators show a larger tuning range and require much smaller electric voltages despite static power consumption. To further improve the modulation response, we could develop thermo-acoustic modulators using suspended LN thin films. Tunable acoustic-wave filters could be developed by integrating microheaters with acoustic cavities or phononic crystals. Utilizing the temperature gradient generated by the microheater on chip, we could develop efficient and compact phase arrays using acoustic waveguides, which have wide applications in beam forming. We believe our thermal control of acoustic waves enable future development of acoustic wave circuits for microwave signal processing.

## Acknowledgment

This work is supported by DOE HEADS-QON grant no. DE-SC0020376, ONR grant no. N00014-20-1-2425, DOD grant no. FA8702-15-D-0001, Raytheon grant no. A40210. L.S. acknowledges funding from Virginia Tech Foundation. N.S. acknowledges funding from the AQT Intelligent Quantum Networks and Technologies (INQNET) research program.

# References

[1] S. Gong, R. Lu, Y. Yang, L. Gao, and A. E. Hassanien, Microwave Acoustic Devices: Recent Advances and Outlook, IEEE Journal of Microwaves **1**, 601 (2021).

[2] P. Delsing *et al.*, The 2019 surface acoustic waves roadmap, J. Phys. D: Appl. Phys. **52**, 353001 (2019).

[3] C. Campbell, *Surface Acoustic Wave Devices and their Signal Processing Applications* (Academic Press, San Diego, CA, 1989).

[4] R. Lu and S. Gong, RF acoustic microsystems based on suspended lithium niobate thin films: advances and outlook, Journal of Micromechanics and Microengineering **31**, 114001 (2021).

[5] K. Yamanouchi, T. Meguro, Y. Wagatsuma, H. Odagawa, and K. Yamamoto, Nanometer Electrode Fabrication Technology Using Anodic Oxidation Resist and Application to Unidirectional Surface Acoustic Wave Transducers, Jpn. J. Appl. Phys. **33**, 3018 (1994).

[6] Y. Takagaki, P. V. Santos, E. Wiebicke, O. Brandt, H. P. Schönherr, and K. H. Ploog, Superhigh-frequency surface-acoustic-wave transducers using AlN layers grown on SiC substrates, Appl. Phys. Lett. **81**, 2538 (2002).

[7] I. V. Kukushkin, J. H. Smet, L. Höppel, U. Waizmann, M. Riek, W. Wegscheider, and K. von Klitzing, Ultrahigh-frequency surface acoustic waves for finite wave-vector spectroscopy of two-dimensional electrons, Appl. Phys. Lett. **85**, 4526 (2004).

[8] Y. Yang, R. Lu, T. Manzaneque, and S. Gong, in *2018 IEEE International Frequency Control Symposium* (IEEE, Olympic Valley, CA, 2018).

[9] Y. Yang, R. Lu, L. Gao, and S. Gong, 10–60-GHz Electromechanical Resonators Using Thin-Film Lithium Niobate, IEEE Trans. Microwave Theory Tech. **68**, 5211 (2020).

[10] S. J. Whiteley, G. Wolfowicz, C. P. Anderson, A. Bourassa, H. Ma, M. Ye, G. Koolstra, K. J. Satzinger, M. V. Holt, F. J. Heremans, A. N. Cleland, D. I. Schuster, G. Galli, and D. D. Awschalom, Spin–phonon interactions in silicon carbide addressed by Gaussian acoustics, Nat. Phys. **15**, 490 (2019).

[11] S. Maity, L. Shao, S. Bogdanović, S. Meesala, Y.-I. Sohn, N. Sinclair, B. Pingault, M. Chalupnik, C. Chia, L. Zheng, K. Lai, and M. Lončar, Coherent acoustic control of a single silicon vacancy spin in diamond, Nat. Commun. **11**, 193 (2020).

[12] S. Maity, B. Pingault, G. Joe, M. Chalupnik, D. Assumpção, E. Cornell, L. Shao, and M. Lončar, Mechanical Control of a Single Nuclear Spin, Phys. Rev. X **12**, 011056 (2022).

[13] Y. Chu, P. Kharel, W. H. Renninger, L. D. Burkhart, L. Frunzio, P. T. Rakich, and R. J. Schoelkopf, Quantum acoustics with superconducting qubits, Science **358**, 199 (2017).

[14] R. Manenti, A. F. Kockum, A. Patterson, T. Behrle, J. Rahamim, G. Tancredi, F. Nori, and P. J. Leek, Circuit quantum acoustodynamics with surface acoustic waves, Nat. Commun. **8**, 975 (2017).

[15] K. J. Satzinger, Y. P. Zhong, H. S. Chang, G. A. Peairs, A. Bienfait, M. H. Chou, A. Y. Cleland, C. R. Conner, E. Dumur, J. Grebel, I. Gutierrez, B. H. November, R. G. Povey, S. J. Whiteley, D. D. Awschalom, D. I. Schuster, and A. N. Cleland, Quantum control of surface acoustic-wave phonons, Nature **563**, 661 (2018).

[16] P. Arrangoiz-Arriola, E. A. Wollack, Z. Wang, M. Pechal, W. Jiang, T. P. McKenna, J. D. Witmer, R. Van Laer, and A. H. Safavi-Naeini, Resolving the energy levels of a nanomechanical oscillator, Nature **571**, 537 (2019).

[17] A. Bienfait, K. J. Satzinger, Y. P. Zhong, H.-S. Chang, M.-H. Chou, C. R. Conner, É. Dumur, J. Grebel, G. A. Peairs, R. G. Povey, and A. N. Cleland, Phonon-mediated quantum state transfer and remote qubit entanglement, Science **364**, 368 (2019).

[18] G. Andersson, B. Suri, L. Guo, T. Aref, and P. Delsing, Non-exponential decay of a giant artificial atom, Nat. Phys. **15**, 1123 (2019).

[19] M. V. Gustafsson, T. Aref, A. F. Kockum, M. K. Ekström, G. Johansson, and P. Delsing, Propagating phonons coupled to an artificial atom, Science **346**, 207 (2014).

[20] M. Mirhosseini, A. Sipahigil, M. Kalaee, and O. Painter, Superconducting qubit to optical photon transduction, Nature **588**, 599 (2020).

[21] W. Jiang, C. J. Sarabalis, Y. D. Dahmani, R. N. Patel, F. M. Mayor, T. P. McKenna, R. Van Laer, and A. H. Safavi-Naeini, Efficient bidirectional piezo-optomechanical transduction between microwave and optical frequency, Nat. Commun. **11**, 1166 (2020).

[22] L. Shao, M. Yu, S. Maity, N. Sinclair, L. Zheng, C. Chia, A. Shams-Ansari, C. Wang, M. Zhang, K. Lai, and M. Lončar, Microwave-to-optical conversion using lithium niobate thin-film acoustic resonators, Optica **6**, 1498 (2019).

[23] B. J. Eggleton, C. G. Poulton, P. T. Rakich, M. J. Steel, and G. Bahl, Brillouin integrated photonics, Nat. Photonics **13**, 664 (2019).

[24] M. Forsch, R. Stockill, A. Wallucks, I. Marinković, C. Gärtner, R. A. Norte, F. van Otten, A. Fiore, K. Srinivasan, and S. Gröblacher, Microwave-to-optics conversion using a mechanical oscillator in its quantum ground state, Nat. Phys. **16**, 69 (2020).
[25] G. S. MacCabe, H. Ren, J. Luo, J. D. Cohen, H. Zhou, A. Sipahigil, M. Mirhosseini, and O. Painter, Nano-acoustic resonator with ultralong phonon lifetime, Science **370**, 840 (2020).
[26] Q. Liu, H. Li, and M. Li, Electromechanical Brillouin scattering in integrated optomechanical waveguides, Optica **6**, 778 (2019).
[27] S. A. Tadesse and M. Li, Sub-optical wavelength acoustic wave modulation of integrated photonic resonators at microwave frequencies, Nat. Commun. **5**, 5402 (2014).
[28] M. B. Assouar, O. Elmazria, P. Kirsch, P. Alnot, V. Mortet, and C. Tiusan, High-frequency surface acoustic wave devices based on AlN/diamond layered structure realized using e-beam lithography, J. Appl. Phys. **101**, 114507 (2007).
[29] L. Fan, X. Sun, C. Xiong, C. Schuck, and H. X. Tang, Aluminum nitride piezo-acousto-photonic crystal nanocavity with high quality factors, Appl. Phys. Lett. **102**, 153507 (2013).
[30] Z. Schaffer and G. Piazza, in *2020 IEEE International Ultrasonics Symposium (IUS)*2020), pp. 1.
[31] Z. A. Schaffer, G. Piazza, S. Mishin, and Y. Oshmyansky, in *2020 IEEE 33rd International Conference on Micro Electro Mechanical Systems (MEMS)*2020), pp. 1281.
[32] J. Wang, M. Park, S. Mertin, T. Pensala, F. Ayazi, and A. Ansari, A Film Bulk Acoustic Resonator Based on Ferroelectric Aluminum Scandium Nitride Films, Journal of Microelectromechanical Systems **29**, 741 (2020).
[33] X.-B. Xu, J.-Q. Wang, Y.-H. Yang, W. Wang, Y.-L. Zhang, B.-Z. Wang, C.-H. Dong, L. Sun, G.-C. Guo, and C.-L. Zou, High-frequency traveling-wave phononic cavity with sub-micron wavelength, arXiv, 2202.07217 (2022).
[34] E. B. Magnusson, B. H. Williams, R. Manenti, M. S. Nam, A. Nersisyan, M. J. Peterer, A. Ardavan, and P. J. Leek, Surface acoustic wave devices on bulk ZnO crystals at low temperature, Appl. Phys. Lett. **106**, 063509 (2015).
[35] T. Kimura, M. Kadota, and Y. Ida, in *2010 IEEE MTT-S International Microwave Symposium*2010), pp. 1740.
[36] T. Takai, H. Iwamoto, Y. Takamine, H. Yamazaki, T. Fuyutsume, H. Kyoya, T. Nakao, H. Kando, M. Hiramoto, T. Toi, M. Koshino, and N. Nakajima, High-Performance SAW Resonator on New Multilayered Substrate Using LiTaO3 Crystal, IEEE Trans Ultrason Ferroelectr Freq Control **64**, 1382 (2017).
[37] M.-H. Li, R. Lu, T. Manzaneque, and S. Gong, Low Phase Noise RF Oscillators Based on Thin-Film Lithium Niobate Acoustic Delay Lines, Journal of Microelectromechanical Systems **29**, 129 (2020).
[38] L. Shao, S. Maity, L. Zheng, L. Wu, A. Shams-Ansari, Y.-I. Sohn, E. Puma, M. N. Gadalla, M. Zhang, C. Wang, E. Hu, K. Lai, and M. Lončar, Phononic Band Structure Engineering for High-Q Gigahertz Surface Acoustic Wave Resonators on Lithium Niobate, Physical Review Applied **12**, 014022 (2019).
[39] C. J. Sarabalis, R. Van Laer, R. N. Patel, Y. D. Dahmani, W. Jiang, F. M. Mayor, and A. H. Safavi-Naeini, Acousto-optic modulation of a wavelength-scale waveguide, Optica **8**, 477 (2021).
[40] F. M. Mayor, W. Jiang, C. J. Sarabalis, T. P. McKenna, J. D. Witmer, and A. H. Safavi-Naeini, Gigahertz Phononic Integrated Circuits on Thin-Film Lithium Niobate on Sapphire, Physical Review Applied **15**, 014039 (2021).
[41] M. Bousquet, P. Perreau, C. Maeder-Pachurka, A. Joulie, F. Delaguillaumie, J. Delprato, G. Enyedi, G. Castellan, C. Eleouet, T. Farjot, C. Billard, and A. Reinhardt, in *2020 IEEE International Ultrasonics Symposium (IUS)*2020), pp. 1.
[42] T. C. Lee, J. T. Lee, M. A. Robert, S. Wang, and T. A. Rabson, Surface acoustic wave applications of lithium niobate thin films, Appl. Phys. Lett. **82**, 191 (2003).
[43] R. Fleury, D. L. Sounas, C. F. Sieck, M. R. Haberman, and A. Alù, Sound Isolation and Giant Linear Nonreciprocity in a Compact Acoustic Circulator, Science **343**, 516 (2014).
[44] Y. Yu, G. Michetti, M. Pirro, A. Kord, D. L. Sounas, Z. Xiao, C. Cassella, A. Alu, and M. Rinaldi, Radio Frequency Magnet-Free Circulators Based on Spatiotemporal Modulation of Surface Acoustic Wave Filters, IEEE Trans. Microwave Theory Tech. **67**, 4773 (2019).
[45] L. Shao, D. Zhu, M. Colangelo, D. H. Lee, N. Sinclair, Y. Hu, P. T. Rakich, K. Lai, K. K. Berggren, and M. Loncar, Electrical control of surface acoustic waves, Nature Electronics **5**, 348 (2022).
[46] H. Mansoorzare and R. Abdolvand, Acoustoelectric Non-Reciprocity in Lithium Niobate-on-Silicon Delay Lines, IEEE Electron Device Letters **41**, 1444 (2020).
[47] L. Bandhu and G. R. Nash, Controlling the properties of surface acoustic waves using graphene, Nano Research **9**, 685 (2016).
[48] R. Sasaki, Y. Nii, Y. Iguchi, and Y. Onose, Nonreciprocal propagation of surface acoustic wave in Ni/LiNbO3, Phys. Rev. B **95**, 020407 (2017).
[49] J. Zhang, B. Peng, Ş. K. Özdemir, Y.-x. Liu, H. Jing, X.-y. Lü, Y.-l. Liu, L. Yang, and F. Nori, Giant nonlinearity via breaking parity-time symmetry: A route to low-threshold phonon diodes, Phys. Rev. B **92**, 115407 (2015).

[50] L. Shao, W. Mao, S. Maity, N. Sinclair, Y. Hu, L. Yang, and M. Lončar, Non-reciprocal transmission of microwave acoustic waves in nonlinear parity–time symmetric resonators, Nature Electronics **3**, 267 (2020).

[51] R. Fleury, D. Sounas, and A. Alu, An invisible acoustic sensor based on parity-time symmetry, Nat. Commun. **6**, 5905 (2015).

[52] D. Hatanaka, I. Mahboob, K. Onomitsu, and H. Yamaguchi, Phonon waveguides for electromechanical circuits, Nat. Nanotechnol. **9**, 520 (2014).

[53] J. C. Taylor, E. Chatterjee, W. F. Kindel, D. Soh, and M. Eichenfield, Reconfigurable quantum phononic circuits via piezo-acoustomechanical interactions, npj Quantum Information **8**, 19 (2022).

[54] R. B. Ward, Temperature coefficients of SAW delay and velocity for Y-cut and rotated $LiNbO_3$, IEEE Transactions on Ultrasonics, Ferroelectrics, and Frequency Control **37**, 481 (1990).